\documentclass{article}
\usepackage[dvips]{color}
\usepackage{amsmath}
\usepackage{amsfonts}   
\usepackage{amssymb}    
\usepackage{multicol}
\usepackage{mathptm,graphicx}
\usepackage{epsf}
\newtheorem{theorem}{Theorem}[section]
\newtheorem{lemma}{Lemma}[section]
\newtheorem{remark}{Remark}[section]
\newtheorem{definition}{Definition}[section]
\newtheorem{example}{Example}[section]
\newtheorem{proposition}{Proposition}[section]

\newtheorem{corollary}{Corollary}[section]
\newcommand{\sce}[1]{\setcounter{equation}{#1}}

\date{}
\begin{document}
\title{Time Consistent Bid-Ask Dynamic
Pricing Mechanisms for Contingent Claims and Its Numerical
Simulations Under Uncertainty
}
\author{Wei Chen \\
Institute of Quantitative Economics\\
 School of Economics\\
    Shandong University\\
    250199, Jinan, China\\
weichen@sdu.edu.cn\\
September 28, 2013 third version\\
April 7, 2013 second version\\
November 17, 2011 first version}
\date{}
\maketitle
\begin{center}
\begin{minipage}{120mm}
\baselineskip 0.2in {\small {\bf Abstract} We study time
consistent dynamic pricing mechanisms of European contingent
claims under uncertainty by using G framework introduced by Peng
([24]). We consider a financial market consisting of a riskless
asset and a risky stock with price process modelled by a geometric
generalized G-Brownian motion, which features the drift
uncertainty and volatility uncertainty of the stock price process.
Using the techniques on G-framework we show that the risk premium
of the asset is uncertain and distributed with maximum
distribution. A time consistent G-expectation is defined by the
viscosity solution of the G-heat equation. Using the time
consistent G-expectation we define the G dynamic pricing mechanism
for the claim. We prove that G dynamic pricing mechanism is the
bid-ask Markovian dynamic pricing mechanism. The full nonlinear
PDE is derived to describe the bid (resp. ask) price process of
the claim. Monotone implicit characteristic finite difference
schemes for the nonlinear PDE are given, nonlinear iterative
schemes are constructed, and the simulations of the bid (resp.
ask) prices of contingent claims under uncertainty are
implemented.
}\\
{\small {\bf Keywords} G Brownian motion, G expectation,
volatility uncertainty, uncertain risk premium, dynamic pricing
mechanism, monotone finite difference}\\
{\small \bf MSC(2000): 60G35, 65M12,91B28}
\end{minipage}
\end{center}
\section{Introduction}
In a complete financial market, it is well known that there exists
unique neutral measure leading unique pricing and hedging for a
given contingent claim, and in an incomplete market it is
impossible that there is unique hedging strategy to hedge a given
contingent claim. With the order processing, inventory, adverse
selection, transaction cost, and illiquid in incomplete market,
research on the bid-ask pricing in the incomplete market is
prevailing. Recently, Madan (see \cite{Madan}, 2012) presents a
two price economies model in a static one period and its
corresponding bid-ask pricing rule, Eberlein, Madan and Pistorius,
etc (see \cite{Eberlein}), give the continuous time bid and ask
price functionals as nonlinear G-expectations (Peng, \cite{PengC})
in two price economies in the context of a Hunt process. Cherny
and Madan (see \cite{ChernyMadan}) develop measures satisfying the
axioms they give and define acceptability indexes in conic
finance, Cherny and Madan (see \cite{Cherny}) make their
applications in finance. Based on time consistent dynamic risk
processes (\cite{Bion1}), Bion-Nadal (see \cite{Bion2}) introduces
an axiomatic approach of time consistent pricing procedure to lead
to the bid-ask dynamic pricing in financial markets with
transaction costs. In this paper, we will work in the G-framework
proposed by Peng (see \cite{PengC}) which is a powerful and
beautiful theoretical analysis tool in the uncertainty economy, to
construct time consistent dynamic bid-ask pricing mechanisms for
the European contingent claims under uncertainty.

In probability framework, the uncertain model of the stock price
is to assume that the stock price be a positive stochastic process
that satisfies the generalized geometric Brownian motion
\begin{eqnarray}\label{eq11}
d \log\tilde{S}_t = \mu_t^p dt +\sigma_t^p dW_t,
\end{eqnarray}
where $(W_t)_{t\ge0}$ be a 1-dimensional standard Brownian motion
defined on a probability space $(W,\mathcal{F},(\mathcal{F}_t
),P)$ and $(\mathcal{F}_t: 0 \leq t \leq T)$ is the filtration
generated by $W_t$, and $(\mu^p_t,\sigma^p_t )_{t\ge 0}$ is
unknown such that
\begin{eqnarray}\label{eq12}
(\mu^p_t,\sigma^p_t) \in [\underline{\mu},\overline{\mu}]\times
[\underline{\sigma},\overline{\sigma}],\underline{\mu},
\overline{\mu}, \underline{\sigma},\overline{\sigma} \mbox{ are
nonnegative constants and }
\underline{\mu}<\overline{\mu},\underline{\sigma}<\overline{\sigma}.
\end{eqnarray}

We denote $\Gamma$ as all possible paths
$(\mu^p_t,\sigma^p_t)_{t\ge 0}$ satisfying ($\ref{eq12}$), then
$\Gamma$ is a closed convex set. For each fixed path $(\gamma_t
)_{t\ge 0} = (\mu^p_t,\sigma^p_t)_{t\ge 0}\in\Gamma$, let
$P_{\gamma}$ be the probability measure on the space of continuous
paths $(C(0,\infty);\mathcal{B}(C(0,\infty)))$ induced by
$\int_0^t(\mu^p_s ds+\sigma^p_s dW_s)$, and denote $P_0$ as the
reference probability measure induced by $W_t$. We set
$\mathcal{P}$ be the class of all such probability measures
$P_{\gamma}$, and for each $P\in \mathcal{P}$ we denote $E_P$ the
corresponding expectation.

If the uncertainty comes from $\mu^p_t$ which is called drift
uncertainty, by Girsanov transform the risky asset price model
could be transfer to risk neutral model. The volatility
uncertainty model was initially studied by Avellaneda, Levy and
Paras \cite{Avell} and Lyons \cite{Lyons} in the risk neutral
probability measures, they intuitively give the bid-ask price of a
European contingent claim as follows
\begin{eqnarray}\label{eq13}
\mbox{ask price:} \sup_{Q\in \mathcal{Q}}
E_Q[\mbox{e}^{-r(T-t)}\xi|\mathcal{F}_t ], \mbox{bid price:}
-\sup_{Q\in\mathcal{Q}} E_Q[-\mbox{e}^{-r(T-t)}\xi|\mathcal{F}_t
],
\end{eqnarray}
where $\mathcal{Q}= \{Q : Q \mbox{ is the risk neutral probability
measure of }P\in \mathcal{P}\}$, $r$ is the short interest rate.

Motivated by the problem of coherent risk measures under the
volatility uncertainty \cite{Artzner2}, Peng (\cite{PengB},
\cite{PengD}) introduced a sublinear expectation on a well defined
sublinear expectation space, under which the canonical process
$(B_t )_{t\ge 0}$ is defined as G-Brownian. The increments of the
G-Brownian motion are zero-mean, independent and stationary and
$N(\{0\}; [\underline{\sigma},\overline{\sigma}])$, and the
corresponding sublinear expectation is called G-expectation. By
using quasi-sure stochastic analysis, Denis, Hu and Peng in
\cite{Denis} construct consistent G-expectation and G-Brownian
motion, and constructed stochastic integral base under G-frame.

Recently, by using G framework, Epstein and Ji \cite{Epstein}
study the utility uncertainty application in economics. For more
general situations, if the uncertainty comes from drift and
volatility coefficients, Peng \cite{Peng3} study the super
evaluation of the contingent claim and utility uncertainty by
using filtration consistent nonlinear expectations theory.

In this paper we study the dynamic pricing mechanism of European
contingent claim written on a risky asset under uncertainty by
using G-framework introduced by Peng (2005, \cite{PengB}). At
first, in a path sublinear space $(\Omega,L_{ip}(\Omega),\hat{E})$
we model the price process of the risky asset by a generalized
G-Brownian motion which describes the drift uncertainty and the
volatility uncertainty of the asset price. Using the techniques in
G-framework we show that the risk premium of the risky asset is
maximum distributed. We define the bid-ask dynamic price of the
claim by using BSDE and derive a bid-ask dynamic pricing formula
by using G-martingale representation theorem \cite{Song}. Further
more we define a time consistent G-expectation $E^G$ by the G-heat
equation and define a corresponding G-Brownian motion. By the
G-expectation $E^G$ transform, the uncertainty model is
transferred to volatility uncertainty model in
$(\Omega,L_{ip}(\Omega),E^G)$, we prove that the condition
G-expectation $E^G_{t,T}$ is the bid-ask dynamic pricing mechanism
for the claim. we also show that the bid-ask dynamic pricing
mechanism $E^G_{t,T}$ is a Markovian dynamic consistent pricing
mechanism and characterize the bid (resp. ask) price by the
viscosity solution of a full nonlinear PDE which is the
Black-Scholes- Barenblatt equation intuitively given by
Avellaneda, Levy and Paras (\cite{Avell}). For numerical computing
the full nonlinear PDE, we propose monotone characteristic finite
difference schemes for discrete solving the nonlinear PDE
equations, provide iterative scheme for the discrete nonlinear
system derived from the characteristic difference discretization,
and analysis the convergence of the iterative solution to the
viscosity solution of the nonlinear PDE. In the end, we give
simulation examples for the ask and bid prices of contingent
claims under uncertainty.

This paper is organized as follows. In Section 2, we give the
financial market model. In Section 3 we derive a bid-ask price
formula for the European contingent claim. In Section 4, we give a
G-martingale transform and G dynamic pricing mechanism for the
claim. Section 5 we investigate the Markovian case for the G
dynamic pricing mechanism and full nonlinear PDEs are derived to
describe the bid and ask prices for the claims. Numerical methods
for the nonlinear PDEs are given in Section 6. The simulation for
the digital option and butterfly option under uncertainty are
shown in Section 7.

\sce{0}
\section{The market model}

We denote by $\Omega = C(R^+)$ the space of all $R-$ valued
continuous paths $(\omega_t )_{t\in R^+} $ with $\omega_0 = 0$,
equipped with the distance
\begin{eqnarray}\label{eq21}
\rho(\omega^1,\omega^2) := \sum_{i=1}^{\infty}2^{-i}[ \max_{t\in
[0,i]}|\omega^1_t-\omega^2_t|\wedge 1],
\end{eqnarray}
then $(\Omega,\rho)$ is a complete separable metric space. For
each fixed $T\in [0,+\infty)$, we denote $\Omega_T
=\{\omega_{\cdot\wedge T}: \omega_t\in \Omega,t\in [0,\infty)\}$.
For the canonical process $(B_t )(\omega) = \omega_t , t \in
[0;+\infty)$, for $\omega\in\Omega$, we set
$$L_{ip}(\Omega_T ):= \{\phi(B_{t_1},\dots,B_{t_n}
) : \forall n\in N, t_1,\dots,t_n \in [0;T], \phi \in
C_{b,Lip}(R^n)\},
$$
and $$L_{ip}(\Omega) := \cup_{n=1}^{\infty}L_{ip}(\Omega_n),$$
where $C_{b,Lip}(R^n)$ denotes the linear space of functions
$\phi$ satisfying
\begin{eqnarray*}
 |\phi(x)-\phi(y)|&\leq& C(1+|x|^m+|y|^m)|x-y| \mbox{ for } x,y\in
 R,\\
&&
\mbox{ some } C
> 0, m\in N\mbox{ is depending on }\phi.
\end{eqnarray*}

Assume that $\underline{\mu}, \overline{\mu}, \underline{\sigma}$
and $\overline{\sigma}$ are nonnegative constants such that
$\underline{\mu}< \overline{\mu}$ and
$\underline{\sigma}<\overline{\sigma}$, we denote
$(\Omega,L_{ip}(\Omega),\hat{E})$ as a sublinear expectation space
such that the canonical process $(B_t)_{t\ge 0}$ is a generalized
G-Brownian motion with
\begin{eqnarray}\label{eq22}
-\hat{E} [-B_t ] =\underline{\mu}t, & \hat{E} [B_t ] =
\overline{\mu}t,\nonumber\\
-\hat{E}[-B^2_t ] =\underline{\sigma}^2t,& \hat{E} [B^2_t ] =
\overline{\sigma}^2t,
\end{eqnarray}
and the generalized G-Brownian motion can be express as follows
\begin{eqnarray}\label{GBM}
B_t = \hat{B}_t +b_t,
\end{eqnarray}
where $(\hat{B}_t )_{t\ge 0}$ is a G-Brownian motion and
$\hat{B}_t$ is $N(\{0\},
[\underline{\sigma}^2t,\overline{\sigma}^2t])$ distributed, and
$b_t$ is $N([\underline{\mu}t,\overline{\mu}t],\{0\})$
distributed. (Peng in \cite{PengC} gave the construction of the
sublinear expectation space $(\Omega,L_{ip}(\Omega), \hat{E} )$,
the generalized G-Brownian motion $(B_t )_{t\ge 0}$, G-Brownian
motion $(\hat{B}_t )_{t\ge 0}$ and
$N([\underline{\mu}t;\overline{\mu}t],\{0\})$ distributed
$(b_t)_{t\ge 0}.)$

In this paper, we consider a financial market with a nonrisky
asset (bond) and a risky asset (stock) continuously trading in
market. The price P(t) of the bond is given by
\begin{eqnarray}\label{eq23}
dP(t) = rP(t)dt, P(0) = 1,
\end{eqnarray}
where $r$ is the short interest rate, we assume a constant
nonnegative short interest rate. The stock price process $S_t$
solves the following SDE
\begin{eqnarray}\label{eq24}
dS_t = S_tdB_t ,
\end{eqnarray}
where $B_t$ is the generalized G-Brownian motion. The properties
showed in $(\ref{eq22})$ imply that the generalized G-Brownian
describe the drift uncertainty and the volatility uncertainty of
the stock price.
\begin{remark}
The generalized G-Brownian motion can be characterized by the
following nonlinear PDE: $u(t;x) = \hat{E}[\phi(x+B_t )]$ is the
viscosity solution of
\begin{eqnarray*}
u(t,x)-g(u(t,x))-G(\partial_{xx}u(t,x)) = 0,& u|_{t=0} = \phi(x),
\end{eqnarray*}
where $\phi(x)$ is a Lipschitz function, $g(\alpha)
=\overline{\mu}\alpha^+-\underline{\mu}\alpha^-$ and $G(\beta)
=\displaystyle\frac{1}{2}(\overline{\sigma}^2\beta^+-\underline{\sigma}^2\beta^-)$
for $\alpha,\beta \in R$.
\end{remark}

For investigation risk premium of the uncertainty model we give
the representation of the process $b_t$ which is
$N([\underline{\mu}t,\overline{\mu}t],\{0\})$ distributed as
follows
\begin{lemma}\label{le21} For each fix $t \in R^+$, we assume that $\mu_t$ be
$N([\underline{\mu},\overline{\mu}],\{0\})$ distributed, and for
$t \ge 0$ we assume that $\mu_t$ is independent from
$(\mu_{t_1},\mu_{t_2},\cdots,\mu_{t_n})$ for each $n\in N$ and
$0\leq t_1,\cdots,t_n< t$. Let $\pi^N_t= \{t^N_0,
t^N_1,\cdots,t^N_N\}$ be a sequence of partitions of $[0, t]$, we
define $\int^t_0\mu_sds =\sum^{N-1}_{k=0}\mu_k(t^N_{k+1}-t^N_k )$,
then $\int_0^t\mu_sds$ is
$N([\underline{\mu}t,\overline{\mu}t],\{0\})$ distributed in
$(\Omega,L_{ip}(\Omega), \hat{E} )$.
\end{lemma}

From Lemma $\ref{le21}$ we have
\begin{lemma}\label{le22}
$<\hat{B}>_t$ is identically distributed with $\int^t_0
\sigma_sds$, where for each fixed $t\in R^+$, $\sigma_t$ is
$N([\underline{\sigma}^2,\overline{\sigma}^2],\{0\})$ distributed,
and for $t\ge 0\ \sigma_t$ is independent from
$(\sigma_{t_1},\sigma_{t_2},\cdots,\sigma_{t_n})$ for each $n\in
N$ and $0\leq t_1,\cdots, t_n \leq t$.
\end{lemma}

From Lemma $\ref{le21}$ we have that $b_t$ is identically
distributed with $\int^t_0\mu_sds$, then the price process of the
stock can be rewritten as follows
\begin{eqnarray}
dS_t=S_t(rdt+\theta_tdt+d\hat{B}_t),
\end{eqnarray}
where $\theta_t$ is risk premium defined by
$$
\theta_t=\mu_t-r.
$$
It is easy to check that $\theta_t$ is
$N([\underline{\mu}-r,\overline{\mu}-r],\{0\})$ distributed.

Consider an investor with wealth $Y_t$ in the market, who can
decide his invest portfolio and consumption at any time $t\in
[0,T]$. We denote $\pi_t$ as the amount of the wealth $Y_t$ to
invest in the stock at time $t$, and $C(t +h)-C(t)\ge 0$ as the
amount of money to withdraw for consumption during the interval
$(t, t +h],h
> 0$. We introduce the cumulative amount of consumption $C_t$ as RCLL
with $C(0) = 0$. We assume that all his decisions can only be
based on the current path information $\Omega_t$.

\begin{definition} A self-financing superstrategy (resp.
substrategy) is a vector process $(Y,\pi,C)$ (resp. $(-Y,\pi,C)$),
where $Y$ is the wealth process, $\pi$ is the portfolio process,
and $C$ is the cumulative consumption process, such that
\begin{eqnarray}
 dY_t =rY_tdt +\pi_td\hat{B}_t +\pi_t\theta_tdt -dC_t,\\
 \mbox{(resp. } -dY_t = -rY_tdt +\pi_td\hat{B}_t+\pi_t\theta_tdt-dC_t\mbox{ )}
 \end{eqnarray}
 where C is an increasing, right-continuous
process with $C_0 = 0$. The superstrategy (resp. substrategy) is
called feasible if the constraint of nonnegative wealth holds
$$
Y_t\geq 0,\ \ t\in[0,T].
$$
\end{definition}
\section{Bid-ask pricing European contingent claim under uncertainty}

From now on we consider a European contingent claim $\xi$ written
on the stock with maturity $T$, here $\xi\in L^2_G (\Omega_T )$ is
nonnegative. We give definitions of superhedging (resp.
subhedging) strategy and ask (resp. bid) price of the claim $\xi$.
\begin{definition}
(1) A superhedging (resp. subhedging) strategy against the
European contingent claim $\xi$ is a feasible self-financing
superstrategy $(Y,\pi,C)$ (resp. substrategy $(-Y,\pi,C)$) such
that $Y_T = \xi$ (resp. $-Y_T =-\xi$). We denote by $\mathcal{H}
(\xi)$ (resp. $\mathcal{H}^{\prime}(-\xi)$) the class of
superhedging (resp. subhedging) strategies against $\xi$, and if
$\mathcal{H} (\xi)$ (resp. $\mathcal{H}^{\prime}(-\xi)$) is
nonempty, $\xi$ is called superhedgeable (resp. subhedgeable).

(2) The ask-price $X(t)$ at time $t$ of the superhedgeable claim
$\xi$ is defined as
$$
X(t)=\inf\{x\ge 0:\exists (Y_t,\pi_t,C_t)\in\mathcal{H}(\xi)\mbox{
such that } Y_t=x\},
$$
and bid-price $X^{\prime}(t)$ at time $t$ of the subhedgeable
claim $\xi$ is defined as
$$
X^{\prime}(t)=\sup\{x\ge 0:\exists
(-Y_t,\pi_t,C_t)\in\mathcal{H}^{\prime}(-\xi)\mbox{ such that }
-Y_t=-x\}.
$$
\end{definition}

Under uncertainty, the market is incomplete and the superhedging
(resp. subhedging) strategy of the claim is not unique. The
definition of the ask-price $X(t)$ implies that the ask-price
$X(t)$ is the minimum amount of risk for the buyer to superhedging
the claim, then it is coherent measure of risk of all
superstrategies against the claim for the buyer. The coherent risk
measure of all superstrategies against the claim can be regard as
the sublinear expectation of the claim, we have the following
representation of bid-ask price of the claim.

\begin{theorem}
Let $\xi\in L^2_G (\Omega_T )$ be a nonnegative European
contingent claim. There exists a superhedging (resp. subhedging)
strategy $(X,\pi,C)\in \mathcal{H}(\xi)$ (resp.
$(-X^{\prime},\pi,C)\in\mathcal{H}^{\prime}(-\xi)$) against $\xi$
such that $X_t$ (resp. $X^{\prime}_t$ ) is the ask (resp. bid)
price of the claim at time $t$.

Let$(H_s^t:s\geq t)$ be the deflator started at time $t$
satisfying
\begin{eqnarray}\label{eq313}
dH_s^t=-H_s^t[rds+\displaystyle\frac{\theta_s}{\sigma_s}d\hat{B}_s],&
H_t^t=1.
\end{eqnarray}
which implies the time value and the uncertain risk value.

Then the ask-price against $\xi$ at time $t$ is
$$
X_t=\hat{E}[H_T^t\xi|\Omega_t],
$$
and the bid-price against $\xi$ at time $t$ is
$$
X^{\prime}_t=-\hat{E}[-H_T^t\xi|\Omega_t].
$$
\end{theorem}
{\bf Proof.} By G-It\^{o}'s formula we can check that
\begin{eqnarray}\label{eq314}
H_t=\exp\{-[\int_0^trds+\int_0^t\displaystyle\frac{\theta_s}{\sigma_s}d\hat{B}_s+\frac{1}{2}\int_0^t(\frac{\theta_s}{\sigma_s})^2d<\hat{B}>_s]\}
\end{eqnarray}
is the solution of ($\ref{eq313}$). Define the stochastic process
$X$ from
$$
H_tX_t=\hat{E}[H_T\xi|\Omega_t]=M_t,
$$
then $M_t$ is a G-martingale. By the G-martingale representation
theorem (\cite{Song}), there exists unique decomposition of
$H_tX_t$ as follows
$$
H_tX_t=\hat{E[H_T\xi]}+\int_0^t\beta_sd\hat{B}_s-K_t,
$$
where $\{\beta_t\}\in H_G^1(0,T)$,$\{K_t\}$ is a continuous,
increasing process with $K_0=0$, and $\{-K_t\}_{0\leq t\leq T}$ is
a G-martingale. Set
$\pi_t=[H_t^{-1}\beta_t+X_t\displaystyle\frac{\theta_t}{\sigma_t}]$.
Then
$H_tX_t=\hat{E}[H_T\xi]+\int_0^tH_s(\pi_s-X_s\displaystyle\frac{\theta_s}{\sigma_s})d\hat{B}_s-K_t$.
Define $C(t)=\int_0^tH_s^{-1}K_sds$, then $C(t)$ is nonnegative
and increasing process with $C(0)=0$. We prove that
$(\hat{E}[H_T^t\xi|\Omega_t],\pi_t,C_t)\in \mathcal{H}(\xi)$ is a
superhedging strategy against $\xi$.

For any superhedging strategy $(Y_t,\hat{\pi}_t,\hat{C}_t)$
against $\xi$, by G-It\^{o}'s formula and
$\sigma_tdt=d<\hat{B}>_t$ which is given by Lemma $\ref{le22}$, we
have
\begin{eqnarray}\label{eq315}
H_tY_t=H_T\xi-\int_0^TH_s(\hat{\pi}_s-Y_s\displaystyle\frac{\theta_s}{\sigma_s})d\hat{B}_s+\int_0^TH_sd\hat{C}_s.
\end{eqnarray}
Taking the condition G-expectatin on both side of $(\ref{eq315})$
with respect to $\Omega_t$, notice that $\hat{C}_t$ is a
nonnegative right continuous process and
$\hat{E}[\int_t^TH_s(\pi_s-Y_s\theta_s)d\hat{B}_s|\Omega_t]=0$, we
have
$$
H_tY(t)\geq\hat{E}[H_T\xi|\Omega_t],
$$
i.e.
$$
Y(t)\geq\hat{E}[H^t_T\xi|\Omega_t]=X_t
$$
which prove that $X_t=\hat{E}[H^t_T\xi|\Omega_t]$ is the ask price
against the claim $\xi$ at time $t$. Similarly we can prove that
$X_t^{\prime}=-\hat{E}[-H_T^t\xi|\Omega_t]$ is the bid price
against the claim $\xi$ at time $t$. $\ \ \square$

\section{G-Girsanov transform and bid-ask dynamic pricing
mechanisms}

In this section, we will construct a time consistent G expectation
$E^G$, and transfer the uncertainty model into the sublinear space
$(\Omega,L_G^2(\Omega),E^G)$ which is correspond with a sequence
of the risk-neutral probability measure space.

Define a sublinear function $G(\cdot)$ as follows
\begin{eqnarray}\label{eq415}
G(\alpha)=\displaystyle\frac{1}{2}(\overline{\sigma}^2\alpha^+-\underline{\sigma}^2\alpha^-),&\forall
\alpha\in R.
\end{eqnarray}
For given $\varphi\in C_{b,lip}(R)$, we denote $u(t,x)$ as the
viscosity solution of the following G-heat equation
\begin{eqnarray}\label{eq416}
\partial_tu-G(\partial_{xx}u)=0,&(t,x)\in(0,\infty)\times R,\\
u(0,x)=\varphi(x).&\nonumber
\end{eqnarray}
\begin{remark}
The G-heat equation ($\ref{eq416}$) is a special kind of
Hamilton-Jacobi-Bellman eqaution, also the Barenblatt equation
except the case $\underline{\sigma} = 0$ (see \cite{Baren1} and
\cite{Baren2}). The existence and uniqueness of ($\ref{eq416}$) in
the sense of viscosity solution can be found in, for example
\cite{Fleming}, \cite{Lions}, and \cite{PengA} for
$C^{1,2}$-solution if $\underline{\sigma} > 0$.
\end{remark}
\begin{theorem}(G-Girsanov transform) Denote
\begin{eqnarray}
\tilde{B}_t=B_t-rt,
\end{eqnarray}
where $(B_t)_{t\geq 0}$ is generalized G-Brownian motion
$(\ref{GBM})$ in sublinear expectation space
$(\Omega,L_{lip}(\Omega),\hat{E})$, then there exists sublinear
expectation $E^G$, such that on the sublinear expectation space
$(\Omega,L_{lip}(\Omega),E^G)$ the process $(\tilde{B}_{t})_{t\geq
0}$ is G-Brownian motion.
\end{theorem}
{\bf Proof.} The stochastic path information of
$\{\tilde{B}_t\}_{t\geq 0}$ up to $t$ is the same as
$\{B_t\}_{t\geq 0}$, without loss of generality we still denote
$\Omega_t$ as the path information of $\{\tilde{B}_t\}_{t\geq 0}$
up to $t$. For $\omega\in\Omega$ consider the process
$(\tilde{B}_t)(\omega)=\omega_t, t\in [0,\infty)$, we define
 $E^G[\cdot]: L_{ip}(\Omega)\longrightarrow R$ as
\begin{eqnarray*}
E^G[\varphi(B_t)]=u(t,0),
\end{eqnarray*}
and for each $s,t\ge 0$ and $t_1,\cdots,t_N\in [0,t]$
\begin{eqnarray*}
E^G[\varphi(\tilde{B}_{t_1},\cdots,\tilde{B}_{t_N},\tilde{B}_{t+s}-\tilde{B}_t)]:=E^G[\psi(\tilde{B}_{t_1},\cdots,\tilde{B}_{t_N})]
\end{eqnarray*}
where
$\psi(x_1,\cdots,x_N)=E^G[\varphi(x_1,\cdots,x_N,\tilde{B}_s)]$.

For $0<t_1<t_2<\cdots<t_{i}<t_{i+1}<\cdots<t_N<+\infty$, we define
G conditional expectation with respect to $\Omega_{t_i}$ as
\begin{eqnarray*}
&&E^G[\varphi(\tilde{B}_{t_1},\tilde{B}_{t_2}-\tilde{B}_{t_1}\cdots,\tilde{B}_{t_{i+1}}-\tilde{B}_{t_{i}},\cdots,
\tilde{B}_{t_N}-\tilde{B}_{t_{N-1}}|\Omega_{t_i}]\\
&:=&\psi(\tilde{B}_{t_1},\tilde{B}_{t_2}-\tilde{B}_{t_1},\cdots,
\tilde{B}_{t_{i}}-\tilde{B}_{t_{i-1}}),
\end{eqnarray*}
where
$\psi(x_1,\cdots,x_i)=E^G[\varphi(x_1,\cdots,x_i,\tilde{B}_{t_{i+1}}-\tilde{B}_{t_{i}},
\cdots,\tilde{B}_{t_N}-\tilde{B}_{t_{N-1}})]$.

We consistently define a sublinear expectation $E^G$ on
$L_{ip}(\Omega)$. Under sublinear expectation $E^G$ we define
above, the corresponding canonical process $(\tilde{B}_t)_{t\ge
0}$ is a G-Brownian motion and $(\tilde{B}_t)$ is
$N(\{0\},[\underline{\sigma}^2t,\overline{\sigma}^2t])$
distributed. $\ \ \ \ \ \ \square$

We call $E^G[\cdot]$ defined in the above proof as G-expectation
on $(\Omega,L_{ip}(\Omega))$. Denote $L_G^p(\Omega),p\geq 1$ as
the completion of $L_{ip}(\Omega)$ under the norm
$\|X\|_p=(E^G[|X|^p])^{1/p}$, and similarly we can define
$L_G^p(\Omega_t)$. The sublinear expectation $E^G[\cdot]$ can be
continuously extended to the space $(\Omega,L_G^1(\Omega))$. From
now on we will work in the sublinear expectation space
$(\Omega,L_G^1(\Omega),E^G)$.

The price dynamic process of the stock ($\ref{eq24}$) can be
rewritten as follows
\begin{eqnarray}\label{418}
dS_t=S_t(rdt+d\tilde{B}_t).
\end{eqnarray}
Denote $D(t):=\mbox{e}^{-rt}$ be the discounted factor, with the
discounted process $\bar{Y}_t=D(t)Y_t,\bar{\pi}_t=D(t)\pi_t$, and
$d\bar{C}_t=D(t)dC_t$, using G-It\^o's formula, we can write the
self-financing superstrategy $(\bar{Y},\bar{\pi},\bar{C})$ (resp.
substrategy $(-\bar{Y},\bar{\pi},\bar{C})$) satisfying
\begin{eqnarray*}
d\bar{Y}_t=\bar{\pi}_td\tilde{B}_t-d\bar{C}\\
 (\mbox{resp.} -d\bar{Y}_t=\bar{\pi}_td\tilde{B}_t-d\bar{C} ).
\end{eqnarray*}
The superhedging (resp. subhedging) strategies and ask (resp. bid)
price of the claim $\xi$ which we defined in Definition 3.1 can
also be characterized using discounted quantities.

For the nonnegative European contingent claim $\xi\in L^2_G
(\Omega_T )$, we define G dynamic pricing mechanism for the claim
$\xi$ as follows:

\begin{definition} For $t \in [0,T]$, we define G dynamic pricing
mechanism as $E^G_{t,T} : L^2_G (\Omega_T )\longrightarrow L^2_G
(\Omega_t )$
$$
 E^G_{t,T} [\cdot] = E^G[\cdot|\Omega_t ].
 $$
\end{definition}

By the comparison theorem of the G-heat equation ($\ref{eq416}$)
and the sublinear property of the function G($\cdot$), similar in
\cite{Peng} the G dynamic pricing mechanisms have the following
properties

\begin{proposition}\label{p41} For $t\in [0,T]$ and $\xi_1,\xi_2\in
L_G^2(\Omega_T)$\\

(i) $E^G_{t,T} [\xi_1] \geq E^G_{t,T} [\xi_2]$ {\mbox if }
$\xi_1\geq
\xi_2$,\\

(ii) $E^G_{T,T} [\xi]=\xi$,\\

(iii) $E^G_{t,T} [\xi_1+\xi_2]\leq E^G_{t,T} [\xi_1]+E^G_{t,T}
[\xi_2]$,\\

(iv) $E^G_{t,T} [\lambda\xi]=\lambda E^G_{t,T} [\xi]$ \mbox{ for } $\lambda \geq 0$,\\

(v) $E^G_{s,t} [E^G_{t,T} [\xi]]=E^G_{s,T} [\xi]$ \mbox{ for }
$0\leq s\leq t$.
\end{proposition}

\begin{theorem}\label{th41}
Assume that $\xi = \phi(S_T)\in L^2_G (\Omega_T )$ be a
nonnegative European contingent claim, and $E^G_{t,T} [\cdot]$ be
the G pricing dynamic mechanisms defined in Definition 4.1. The
ask price and bid price against the contingent claim $\xi$ at time
t are
\begin{eqnarray}
u^a(t,S_t) = \mbox{e}^{-r(T-t)}E^G_{t,T} [\xi] \mbox{ and }
u^b(t,S_t ) = -\mbox{e}^{-r(T-t)}E^G_{t,T} [-\xi],
\end{eqnarray} respectively.
\end{theorem}
{\bf Proof.} It is easy to check that $\{u_t^{a}\}_{0<t<T}$
satisfying
$$
D(t)u_t^a=E_{t,T}^G[D(T)\xi].
$$
Then $M_t=D(t)u_t^a$ is a G-martingale. By a similar way we used
in the proof of Theorem 3.1, we can complete the proof. $\ \
\square$
\begin{remark}
In the Proposition $\ref{p41}$, (iii) and (iv) imply that
$$
E_{t,T}^G[\alpha\xi_1+(1-\alpha)\xi_2]\leq \alpha
E_{t,T}^G[\xi_1]+(1-\alpha)E_{t,T}^G[\xi_2]\ \ \mbox{for}\
\alpha\in[0,1],
$$
which means that the G dynamic pricing mechanism is a convex
pricing mechanism. The equality (v) means the G dynamic pricing
mechanism is a time consistent markivian pricing mechanism, under
the G dynamic pricing mechanism the ask price
$u^a(s;S_s)=\mbox{e}^{-r(T-s)}E^G_{s,T} [\xi]$ (resp. the bid
price $u^b(s,S_s)=-\mbox{e}^{-r(T-s)}E^G_{s,T} [-\xi]$) at time $s
(s\leq t\leq T)$ against the claim $\xi$ with maturity $T$ could
be regarded as the ask (resp. bid) price at time $s$ against the
claim $u^a(t,S_t) = \mbox{e}^{-r(T-t)}E^G_{t,T} [\xi]$ (resp.
$u^b(t,S_t ) = -\mbox{e}^{-r(T-t)}E^G_{t,T} [-\xi]$) with maturity
$t$.
\end{remark}

\section{Markovian case}

We assume that the stock price dynamic process satisfying the
following SDE ($t\geq 0$)
\begin{eqnarray}\label{eq520}
d S_s^{t,x}&=&S_s^{t,x}[r dt + d\tilde{B}_t],\ \  s\in [t,T],\\
S_t^{t,x}&=&x.\nonumber
\end{eqnarray}

For given a nonnegative European contingent claim $\xi = \phi(S_T
)\in L^2_G (\Omega)$, from Theorem $\ref{th41}$ its ask price and
bid price at time $t$ are
\begin{eqnarray*}
u^a(t,x) = \mbox{e}^{-r(T-t)}E^G_{t,T} [\phi(S_T^{t,x})],\\
 u^b(t,x )
= -\mbox{e}^{-r(T-t)}E^G_{t,T} [-\phi(S_T^{t,x})].
\end{eqnarray*}

We establish the ask (resp. bid) price as the viscosity solution
of a full nonlinear PDE as follows:

\begin{theorem}\label{th51} Assume that the stock price dynamic process
satisfying ($\ref{eq520}$), $\xi = \phi(S^{t,x}_T)\in L^2_G
(\Omega_T )$ be a nonnegative European contingent claim written on
the stock with maturity $T$, and $\phi: R \longrightarrow R$ be a
given Lipschitz function. The ask price of the contingent claim
$u^a(t,x) = \mbox{e}^{-r(T-t)}E^G_{t,T} [\xi]$ is the viscosity
solution of the following nonlinear PDE
\begin{eqnarray}\label{eq521}
\partial_t u^a(t,x)+rx\partial_x u^a(t,x)+G(x^2\partial_{xx}u^a(t,x))-ru^a(t,x) = 0,& (t,x)
\in [0,T)\times R,&\\
u^a(T,x) = \phi(x).\nonumber
 \end{eqnarray}
  The bid price of the contingent
claim $u^b(t,x) =-\mbox{e}^{-r(T-t)}E^G_{t,T} [-\xi]$ is the
viscosity solution of the following nonlinear PDE
\begin{eqnarray}\label{eq522}
\partial_t u^b(t,x)+rx\partial_x u^b(t,x)-G(-x^2\partial_{xx}u^b(t,x))-ru^b(t,x) = 0,& (t,x)
\in [0,T)\times R,&\\
u^b(T,x) = \phi(x).\nonumber
 \end{eqnarray}
\end{theorem}
{\bf Proof.} With the assumption that $\xi\in L_G^2(\Omega_T)$,
Peng in \cite{PengC} prove that
$v(t,x)=E_{t,T}^G[\phi(S_T^{t,x})]$ satisfying
$$
|v(t,x)-v(t,x')|\leq C|x-x'|,\ \ |v(t,x)|\leq C(1+|x|),
$$
and for $\delta\in[0,T-t]$
$$
|v(t,x)-v(t+\delta,x)|\leq C(1+|x|)(\delta^{1/2}+\delta),\ \
\delta\in [0,T-t],
$$
$$
v(t,x)=E^G[v(t+\delta,S_{t+\delta}^{t,x})],
$$
where $C$ is only dependent on the Lipschitz constant.

We can easily get that
\begin{eqnarray*}
|u^a(t,x)-u^a(t,x')|\leq C|x-x'|,\ \ |u^a(t,x)|\leq C(1+|x|),\\
|u^a(t,x)-u^a(t+\delta,x)|\leq C(1+|x|)(\delta^{1/2}+\delta),
\end{eqnarray*}
and
$$
u^a(t,x)=E^G[e^{-r\delta}u^a(t+\delta,S_{t+\delta}^{t,x})].
$$

For fixed $(t,x)\in (0,T)\times R$, let $\psi\in
C_b^{2,3}([0,T]\times R)$ be such that $\psi\ge u^a$ and
$\psi(t,x)=u^a(t,x)$. By Taylor's expansion, we have for
$\delta\in (0, T-t)$
\begin{eqnarray*}
0&\leq& E^G[\mbox{e}^{-r\delta}\psi(t+\delta,
S_{t+\delta}^{t,x})-\psi(t,x)]\\
&\leq
&\displaystyle\frac{1}{2}E^G[x^2\partial_{xx}\psi(t,x)(<B>_{t+\delta})-<B>_{t})]\\
&&+(\partial_t \psi(t,x)+rx\partial_x
\psi(t,x)-r\psi(t,x))\delta+C(1+|x|+|x|^2+|x|^3)\delta^{3/2}\\
&\leq&(\partial_t \psi(t,x)+rx\partial_x
\psi(t,x)+G(x^2\partial_{xx}\psi(t,x))-r\psi(t,x))\delta\\
&&+C(1+|x|+|x|^2+|x|^3)\delta^{3/2},
\end{eqnarray*}
for $\delta\downarrow 0$, we have
$$
\partial_t \psi(t,x)+rx\partial_x
\psi(t,x)+G(x^2\partial_{xx}\psi(t,x))-r\psi(t,x)\geq 0,
$$
which implies that $u^a(t,x)$ is the subsolution of the nonlinear
PDE ($\ref{eq521}$), and by a similar way we can prove that
$u^a(t,x)$ is the supersolution of ($\ref{eq521}$). Thus, we prove
that the ask price against the claim $\xi$ at time $t$ is the
viscosity solution of ($\ref{eq521}$). Similarly, we can prove
that $u^b(t,x)$ is the viscosity solution of ($\ref{eq522}$). $\ \
 \square$

Assume that the stock price solve the following SDE
\begin{eqnarray}\label{eq523}
d \log\tilde{S}_{u}^{t,x}& =& rdt + \tilde{\sigma}_tdW_t,\ \  u\in
[t,T],\\
\tilde{S}_{u}^{t,x}&=&x,\nonumber
 \end{eqnarray}
where $(W_t )_{t\geq 0}$ be a 1-dimensional standard Brownian
motion defined on a probability space $(\Omega,\mathcal{F},P)$ and
the filtration generated by $W_t$ is ($\mathcal{F}_t : 0 \leq t
\leq T)$. Assume that $(\tilde{\sigma}_t )_{t\geq 0}$ is an
adapted process such that $\tilde{\sigma}_t\in
[\underline{\sigma},\overline{\sigma}]$. It is well known that the
Black-Scholes type price against the European contingent claim
$\xi = \phi(\tilde{S}_T^{t,x})$ at time $t$ is $u(t,x) =
\mbox{e}^{-r(T-t)}E[\varphi(\tilde{S}_T^{t,x}|\mathcal{F}_t ]$
which is the viscosity solution of the following PDE
\begin{eqnarray}\label{eq524}
\partial_t u(t,x)+rx\partial_x u(t,x)+\frac{1}{2}x^2\tilde{\sigma}_t^2\partial_{xx}u(t,x))-ru(t,x) = 0,& (t,x)
\in [0,T)\times R,&\\
u(T,x) = \phi(x).\nonumber
\end{eqnarray}

\begin{corollary} Assume that $\phi:R\longrightarrow R$ is a
given Lipschitz function,$u(t,x)$ is the Black-Scholes type price
against the claim $\phi(\tilde{S}_{T}^{t,x})$ which satisfies
($\ref{eq523}$). Assume that $u^a(t,x)$ and $u^b(t,x)$ satisfying
($\ref{eq522}$) and ($\ref{eq521}$), i.e., be the bid price and
ask price against the European contingent claim
$\xi=\phi(S_T^{t,x})$ at time $t$. Then we have
\begin{eqnarray}
u^b(t,x)\leq u(t,x)\leq u^a(t,x)
\end{eqnarray}
\end{corollary}
{\bf Proof.} By using the comparison theorem proposed by Peng in
\cite{PengC}, we can easily prove the Corollary. $\ \ \square$
\begin{lemma}\label{le51}
Assume that the price process $(S_s^{t,x})_{s>t}$ of the stock
satisfies ($\ref{eq520}$), $\xi=\phi(S_T^{t,x})\in
L_G^2(\Omega_T)$ be a nonnegative European contingent claim
written on stock with maturity $T$, and $\phi: R\longrightarrow R$
be a given Lipschitz function.

(I) If $\phi(\cdot)$ is convex (resp. concave), for any $t\in
[0,T]$ the ask price function $u^a(t,\cdot)$ is convex (resp.
concave), and $u^a(t,x)$ satisfies ($\ref{eq521}$) with
$G(x^2\partial_{xx}u^a)=\frac{1}{2}x^2\overline{\sigma}_t^2\partial_{xx}u^a(t,x)$
(resp.
$=\frac{1}{2}x^2\underline{\sigma}_t^2\partial_{xx}u^a(t,x)$). If
$\phi(\cdot)$ is convex (resp. concave) on interval $(a,b)\in R\
(a<b)$ the ask price function $u^a(t,\cdot)$ is convex (resp.
concave) on $(a,b)$ for any $t\in [0,T]$.

(II) If $\phi(\cdot)$ is convex (resp. concave), for any
$t\in[0,T]$ the bid price function $u^b(t,\cdot)$ is convex (resp.
concave), and $u^b(t,x)$ satisfies ($\ref{eq522}$) with
$G(-x^2\partial_{xx}u^b)=-\frac{1}{2}\underline{\sigma}^2x^2\partial_{xx}u^b$
(resp.$=-\frac{1}{2}\overline{\sigma}^2x^2\partial_{xx}u^b$). If
$\phi(\cdot)$ is convex (resp. concave) on interval $(a,b)\in R\
(a<b)$ the bid price function $u^b(t,x)$ is convex (resp. concave)
on $(a,b)$ for any $t\in [0,T]$.
\end{lemma}
{\bf Proof.} We only prove (I).

1. First we prove that $u^a(t,x)$ is convex if $\phi(x)$ is
convex.

For $x_1,x_2\in R$ and for given $\alpha\in[0,1]$, by G-It\^{o}'s
formula, we get the stock price process as follows
\begin{eqnarray}\label{eq526}
&&S_T^{t,\alpha x_1+(1-\alpha )x_2}\nonumber\\
&=& (\alpha x_1+(1-\alpha
)x_2)\exp(r(T-t)+\tilde{B}_T-\tilde{B}_t-\displaystyle\frac{1}{2}(<\tilde{B}>_T-<\tilde{B}>_t))\nonumber\\
&=&\alpha x_1\exp(r(T-t))+\tilde{B}_T-\tilde{B}_t-\displaystyle\frac{1}{2}(<\tilde{B}>_T-<\tilde{B}>_t))\nonumber\\
&&+(1-\alpha) x_2\exp(r(T-t))+\tilde{B}_T-\tilde{B}_t-\displaystyle\frac{1}{2}(<\tilde{B}>_T-<\tilde{B}>_t))\nonumber\\
&=&\alpha S_T^{t,x_1}+(1-\alpha) S_T^{t,x_2}
\end{eqnarray}
Since the G pricing dynamic mechanism is a convex pricing
mechanism, we have that
\begin{eqnarray}\label{eq527}
&&u^a(t,\alpha x_1+(1-\alpha )x_2)\nonumber\\
&=& e^{-r(T-t)}E_{t,T}^G[\phi(S_T^{t,\alpha x_1+(1-\alpha
)x_2})]\nonumber\\
&=& e^{-r(T-t)}E_{t,T}^G[\phi(\alpha S_T^{t,x_1}+(1-\alpha
)S_T^{t,x_2})]\nonumber\\
&\leq &e^{-r(T-t)}E_{t,T}^G[\alpha\phi(S_T^{t,x_1})+(1-\alpha
)S_T^{t,x_2}]\nonumber\\
&\leq & \alpha e^{-r(T-t)}E_{t,T}^G[\phi(S_T^{t,x_1})]+(1-\alpha
)e^{-r(T-t)}E_{t,T}^G[S_T^{t,x_2}]\nonumber\\
&=&\alpha u^a(t,x_1)+(1-\alpha)u^a(t,x_2)
\end{eqnarray}
which proves that $u^a(t,x)$ is convex on $R,\partial_{xx}u^a$ is
nonnegative on $R$ and
$G(x^2\partial_{xx}u^a)=\frac{1}{2}\overline{\sigma}^2x^2\partial_{xx}u^a$.

If $\phi(x)$ is convex in interval $(a,b)$, for $x_1,x_2\in
(a,b)$, ($\ref{eq526}$), ($\ref{eq527}$) hold on $(a,b)$ and
$u^a(t,x)$ is convex on $(a,b)$.

2. We will prove that $u^a(t,x)$ is concave if $\phi(x)$is
concave.

For given process $(\sigma_t)_{t\geq 0}$ such that $\sigma_t\in
[\underline{\sigma}_t,\overline{\sigma}_t]$, we denote
$u_{\sigma}^a$ as the viscosity solution of the following PDE
\begin{eqnarray*}
\partial_t u^a_{\sigma}(t,x)+\frac{1}{2}x^2\sigma_t^2\partial_{xx}u^a_{\sigma}(t,x)&=& 0, (t,x)
\in R^+\times R,\\
u^a_{\sigma}(T,x)& =& \phi(x).\nonumber
\end{eqnarray*}
then
$u^a_{\sigma}=E_t[\phi(x\exp[\sigma_t(W_T-W_t)-\frac{1}{2}\sigma_t^2(T-t)])]$,
where $W_t$ is the standard Brownian motion defined on a
probability space $(\Omega,\mathcal{F},(\mathcal{F}_t),P)$, and
$E_t$ is the corresponding condition expectation. Denote
$S_{T,\sigma}^{t,x}=x\exp[\sigma_t(W_T-W_t)-\frac{1}{2}\sigma_t^2(T-t)]$,
for $x_1,x_2\in R$ if $\phi(x)$ is concave on $R$, we have
\begin{eqnarray}\label{eq528}
&&u^a_{\sigma}(t,\alpha x_1+(1-\alpha )x_2)\nonumber\\
&=&E_{t}[\phi((\alpha x_1+(1-\alpha
)x_2)\exp[\sigma_t(W_T-W_t)-\frac{1}{2}\sigma_t^2(T-t)])]
\nonumber\\
&=&E_{t}[\phi(\alpha S_{T,\sigma}^{t,x_1}+(1-\alpha
)S_{T,\sigma}^{t,x_2})]\nonumber\\
&\geq& E_{t}[\alpha\phi( S_{T,\sigma}^{t,x_1})+(1-\alpha
)\phi(S_{T,\sigma}^{t,x_2})]\nonumber\\
&=&\alpha E_{t}[\phi( S_{T,\sigma}^{t,x_1})]+(1-\alpha
)E_t[\phi(S_{T,\sigma}^{t,x_2})]\nonumber\\
&=&\alpha u_{\sigma}^a(t,x_1)+(1-\alpha ) u_{\sigma}^a(t,x_2)
\end{eqnarray}
which mean $u_{\sigma}^a(t,\cdot)$ is concave on $R$, and the
function $U(t,x)=\sup_{\sigma\in
[\underline{\sigma},\overline{\sigma}]}u_{\sigma}^a(t,x)$ is
concave on $R$, i.e.
\begin{eqnarray}\label{eq529}
U(t,\alpha x_1+(1-\alpha)x_2)\geq \alpha U(t,x_1)+ (1-\alpha)
U(t,x_2).
\end{eqnarray}
Since the operator $\sup_{\sigma\in
[\underline{\sigma},\overline{\sigma}]}E_t[\cdot]$ is a convex
operator and $U(t,x)=\sup_{\sigma\in
[\underline{\sigma},\overline{\sigma}]}E_t[\phi(S_{T,\sigma}^{t,x})]$,
using the similar argument in Theorem $\ref{th51}$, we can prove
that $U(t,x)$ is the viscosity solution of the following HJB
equation
\begin{eqnarray}\label{eq530}
\partial_t U(t,x)+G(x^2\partial_{xx}U(t,x))&=&0,\
\
(t,x)\in R^+\times R,\\
U(T,x)&=&\phi(x),\nonumber
\end{eqnarray}
In \cite{PengC}, Peng prove that
$U(t,x)=E_{t,T}^G[\phi(S_T^{t,x})]$ is the unique viscosity
solution of ($\ref{eq530}$), from ($\ref{eq529}$) the solution
$U(t,x)$ is concave for $x$ on $R$. Thus we prove that
$u^a(t,x)=e^{-r(T-t)}U(t,x)$ is concave for $x$ on $R$ and
$G(x^2\partial_{xx}u^2)=\frac{1}{2}\underline{\sigma}x^2\partial_{xx}u^a$.

If $\phi(x)$ is concave on $(a,b)$, ($\ref{eq528}$) and
($\ref{eq529}$) hold on $(a,b)$, which prove that $u^a(t,x)$ is
concave on $(a,b)$.

We finish the proof of (I).$\ \ \square$

\section{Monotone characteristic finite difference schemes}
In this section we will consider numerical schemes for the
nonlinear PDE ($\ref{eq521}$) (similar for ($\ref{eq522}$)).
\subsection{Characteristic finite difference schemes}
Define $u(\tau,x)=u^a(T-t,x)$, the nonlinear PDE ($\ref{eq521}$)
(similar for ($\ref{eq522}$)) can be written as
\begin{eqnarray}\label{HJB11}
\partial_{\tau}u(\tau,x)-rx\partial_{x}u(\tau,x)-x^2 G(\partial_{xx}u(\tau,x))+ru(\tau,x)&=&
0,\ \ (\tau,x)\in(0,T]\times R,\nonumber\\
u(0,x)&=&\phi(x).
\end{eqnarray}

First, we consider the boundary conditions of ($\ref{HJB11}$). As
$x\longrightarrow 0$, equation ($\ref{HJB11}$) becomes
\begin{eqnarray}\label{eq62}
\partial_{\tau}u|_{x=0}=-ru(\tau,0).
\end{eqnarray}

For $x\longrightarrow \infty$, normally for sufficient big enough
$x$ there holds $\partial_{xx}u \simeq 0$. We set $x=S_{max}$ with
$S_{max}$ big enough price of the stock which makes the payoff has
asymptotic form. We consider the Dirichlet condition as follows
\begin{eqnarray}\label{eq63}
u|_{x=S_{max}}=g(\tau,S_{max}),
\end{eqnarray}
where $g(.,l)$ can be determined by financial reasoning for some
given contingent claim, normally in the following asymptotic form
$$
g(\tau,S_{max})=b(\tau)S_{max}+c(\tau).
$$

We assume that $b(\tau),c(\tau)$ are bounded such that
\begin{eqnarray}
|u(\tau,S_{max})|\leq C_b,
\end{eqnarray}
where $C_b$ is a constant.

There is convection term
$\partial_{\tau}u(\tau,x)-rx\partial_{x}u(\tau,x)$ in
$(\ref{HJB11})$, if the convection dominate the diffusion the
finite difference discretization for $(\ref{HJB11})$ could leads
numerical oscillations, we consider discrete the convection term
along the characteristic direction (\cite{Douglas}). Denote $
\psi(s)=[1+r^2x^2]^{1/2}$, the direction derivative along
characteristic direction is $\frac{\partial}{\partial
c}=\frac{1}{\psi}(\frac{\partial}{\partial
\tau}-rx\frac{\partial}{\partial x})$.
($\ref{HJB11}$)-($\ref{eq63}$) are equivalent to the following
equation
\begin{eqnarray}\label{HJB12}
\psi(x)\frac{\partial u}{\partial c}-\sup_{\underline{\sigma} \leq
\sigma\leq\overline{\sigma}}\mathcal{L}^{\sigma}u=
0,&&\ \ (\tau,x)\in(0,T]\times R,\nonumber\\
\partial_{\tau}u|_{x=0}=-ru(\tau,0),\ \ u|_{x=S_{max}}=g(\tau,S_{max}),&&\nonumber\\
u(0,x)=\phi(x),&&
\end{eqnarray}
where
$\mathcal{L}^{\sigma}u=\frac{\sigma^2}{2}x^2\partial_{xx}u-ru$.

Now we define spatial partition of $I=[0,S_{max}]$. Let
$I=[0,S_{max}]$ be divided into $N$ sub-intervals
$$
I_i=(x_i,x_{i+1}),\ \ i=0,...,N-1.
$$
with $0=x_0<x_1<\cdots<x_N=S_{max}$. For each $i=0,...,N-1$, let
$\Delta x_i=x_{i+1}-x_i$. Let $t_i\ (i=0,1,\cdots,M)$ be a set of
partition point in $[0,T]$ satisfying $0=t_0<t_1<\cdots<t_M=T$,
and denote $\Delta t_n=t_n-t_{n-1}>0$, where $M>1$ is a positive
integer.

Let $u_i^n$ be a discrete approximation to $u(t_n,x_i)$. Denote
$\bar{x}_{i}^n=x_i+rx_i\Delta t_{n+1}$, for $\Delta t_{n+1}$ small
enough such that $\bar{x}_i^n\in [x_i,x_{i+1}]$. Denote
$\bar{u}_i^n$ be a discrete approximation to
$u(t_n,\bar{x}_i^{n})$, here we define $\bar{u}_i^n$ as the linear
interpolate function of $u_i^n$. The implicit characteristic
finite difference scheme for ($\ref{HJB11}$) is as follows:

For boundary $x=0$
\begin{eqnarray}\label{eq66}
\displaystyle\frac{u_0^{n+1}-u_0^{n}}{\Delta t_{n+1}}=-ru_0^{n+1},
\end{eqnarray}
for $i=1,2,\cdots,N-1$

\begin{eqnarray}\label{eq67}
\frac{u_i^{n+1}-u_i^{n}}{\Delta
t_{n+1}}=\sup_{\sigma^{n+1}\in\{\underline{\sigma},\overline{\sigma}\}}[(L_h^{\sigma^{n+1}}u^{n+1})_i]
+ \frac{\bar{u}_i^{n}-u_i^{n}}{\Delta t_{n+1}},
\end{eqnarray}
where $(L_h^{\sigma^{n}}u^{n})_i$ denotes the central difference
discrete of $(\L^{\sigma} u)$ at $(t_n,x_i)$, i.e.
\begin{eqnarray}\label{eq68}
(L_h^{\sigma^{n}}u^{n})_i=\alpha_{i}^{n}(\sigma_i^{n})u_{i-1}^{n}
+\beta_{i}^{n}(\sigma_i^{n})u_{i+1}^{n}
-(\alpha_{i}^{n}(\sigma_i^{n})
+\beta_{i}^{n}(\sigma_i^{n})+r)u_{i}^{n},
\end{eqnarray}
where $\alpha_i^n$ and $\beta_i^n$ are defined as follows
\begin{eqnarray}\label{eq69}
\begin{array}{r}
\alpha_{i}^{n}(\sigma_i^{n})=\displaystyle\frac{(\sigma_i^{n})^2x_i^2}{(x_i-x_{i-1})(x_{i+1}-x_{i-1})},\\
\beta_{i}^{n}(\sigma_i^{n})=\displaystyle\frac{(\sigma_i^{n})^2x_i^2}{(x_{i+1}-x_{i})(x_{i+1}-x_{i-1})}.
\end{array}
\end{eqnarray}
It is easy to check that $\alpha_i^n,\beta_i^n\geq 0$. We define
\begin{eqnarray*}
U^n=[u_0^n,\cdots,u_{N-1}^n,u_{N}^n]^T,\\
\bar{U}^n=[\bar{u}_0^n,\cdots,\bar{u}_{N-1}^n,\bar{u}_{N}^n]^T,\\
\sigma^n=[\sigma_1^n,\cdots,\sigma_{N-1}^n]^T,\\
\end{eqnarray*}
and
$$
(A^n(\sigma^n)U^n)_i=\alpha_i^n(\sigma_i^n)u_{i-1}^n+\beta_i^n(\sigma_i^n)u_{i+1}^n-(\alpha_i^n(\sigma_i^n)+\beta_i^n(\sigma_i^n)+r)u_{i}^n.
$$
For notational consistency, we denote
$\bar{u}_0^n=\displaystyle\frac{1}{1+r\Delta t
_{n+1}}u_0^n$,$\bar{u}_N^n=u_N^{n+1}$, and enforce the first row
and the last row of A to be zero. Then the discretization scheme
($\ref{eq67}$) can be write as the following equivalent matrix
form
\begin{eqnarray}\label{eq610}
\left\{
\begin{array}{c}
[I-\Delta t_{n+1}\hat{A}^{n+1}]U^{n+1}=\bar{U}^n\\
\hat{\sigma}_i^{n+1}=\mbox{argsup}_{\sigma^{n+1}\in
\{\underline{\sigma},\overline{\sigma}\}}\{(A^{n+1}U^{n+1})_i\}
\end{array}
\right.
\end{eqnarray}
where $\hat{A}^{n+1}=A(\hat{\sigma}^{n+1})$ and
$\hat{\sigma}^{n+1}=\hat{\sigma}^{n+1}(U^{n+1})$.

It is easy to see that $-\hat{A}^{n+1}$ has nonpositive
off-diagonals, positive diagonal, and is diagonally dominate. We
have the following theorem:
\begin{theorem}\label{th61}
Matrices $-\hat{A}^{n+1}$ and $I-\Delta t_{n+1}\hat{A}^{n+1}$ are
M-matrices (\cite{Varga}).
\end{theorem}
The equation ($\ref{HJB11}$) has unique viscosity solution, and
satisfies the strong comparison property (see \cite{Chaumont} and
\cite{Lions}), then a numerical scheme converges to the viscosity
solution if the method is consistent, stable and monotone.

Let $h=\max\{\Delta x,\Delta t\}$ be the mesh parameter, where
$\Delta x=\max_i\Delta x_i,\Delta t=\max_n\Delta t_n$. Assume that
the partition is quasi-uniform, i.e., $\exists C_1,C_2>0$
independent of $h$ such that
\begin{eqnarray}\label{eq611}
C_1h\leq \Delta x_i,\Delta t_n\leq C_2 h
\end{eqnarray}
for $0\leq i\leq N-1$ and $1\leq n\leq M$.

We denote

\begin{eqnarray}\label{eq612}
G_i^{n+1}(h,u_{i-1}^{n+1},u_{i}^{n+1},u_{i+1}^{n+1},\bar{u}_i^{n})&=&
\frac{u_i^{n+1}-u_i^{n}}{\Delta
t_{n+1}}-\sup_{\sigma^{n+1}\in\{\underline{\sigma},\overline{\sigma}\}}(A^{n+1}U^{n+1})_i
 - \frac{\bar{u}_i^{n}-u_i^{n}}{\Delta t_{n+1}}.
\label{GO}
\end{eqnarray}
Then the discrete equation at each node can be written as the
following form
\begin{eqnarray}
G_i^{n+1}(h,u_{i-1}^{n+1},u_{i}^{n+1},u_{i+1}^{n+1},\bar{u}_i^{n})=0.
\end{eqnarray}
\begin{lemma}(Stability)\label{le61}
The discretizaition ($\ref{eq610}$) is stable i.e.
\begin{eqnarray}
\|U^n\|_{\infty}\leq \max(\|U^0\|_{\infty},C_b).\label{stable}
\end{eqnarray}
\end{lemma}
{\bf Proof.} The discrete equations are
\begin{eqnarray*}
(1+\Delta
t_{n+1}(\alpha_i^{n+1}+\beta_i^{n+1}+r))u_i^{n+1}=\bar{u}_i^{n}+\Delta
t_{n+1}\alpha_i^{n+1}u_{i-1}^{n+1}+\Delta
t_{n+1}\beta_i^{n+1}u_{i+1}^{n+1}.
\end{eqnarray*}
Since we take $\bar{u}_i^n$ as linear interpolation of $u_i^n$ and
$u_{i+1}^n$ in the discretization, we obtain
\begin{eqnarray*}
(1+\Delta t_{n+1}(\alpha_i^{n+1}+\beta_i^{n+1}+r))|u_i^{n+1}|\leq
\|U^n\|_{\infty}+\Delta
t_{n+1}(\alpha_i^{n+1}+\beta_i^{n+1})\|U^{n+1}\|_{\infty}.
\end{eqnarray*}

If $\|U^{n+1}\|_{\infty}=|u_{j}^{n+1}|,\ 1<j<N$, then we have
\begin{eqnarray*}
(1+\Delta
t_{n+1}(\alpha_i^{n+1}+\beta_i^{n+1}+r))\|U^{n+1}\|_{\infty}\leq
\|U^n\|_{\infty}+\Delta
t_{n+1}(\alpha_j^{n+1}+\beta_j^{n+1})\|U^{n+1}\|_{\infty}
\end{eqnarray*}
which implies that
\begin{eqnarray*}
\|U^{n+1}\|_{\infty}\leq \|U^{n}\|_{\infty}.
\end{eqnarray*}
If $j=0$ or $j=N$, then $\|U^{n+1}\|_{\infty}=|u_0^{n+1}|\leq
|u_0^n|$ or $\|U^{n+1}\|_{\infty}=|u_N^{n+1}|\leq C_b$.

Thus, we have
\begin{eqnarray*}
\|U^{n+1}\|_{\infty}\leq \max(\|U^0\|_{\infty},C_b)
\end{eqnarray*}
which complete the prove.$\ \ \square$

\begin{lemma}(Consistency) For any smooth function $v$ with $v_i^n=v(t^n,x_i)$, the discrete scheme ($\ref{eq610}$) is consistent.
\end{lemma}
{\bf Proof.} Using Taylor series expansions, we can have
\begin{eqnarray*}
|(\frac{1}{2}\sigma\partial_{xx}v-rv)_i^n-(L_h^{\sigma^n}v^{n})_i|=O(\Delta
x)
\end{eqnarray*}
and using expansion along characteristic direction $c$
\begin{eqnarray*}
&&|(\psi\frac{\partial v}{\partial c})_i^{n+1}-\frac{v_i^{n+1}-\bar{v}_i^{n}}{\Delta t_{n+1}}|\nonumber\\
&=& O(\Delta x+\Delta t),
\end{eqnarray*}
where $\bar{v}_i^n=v(t^n,\bar{x}_i^n)$.

For smooth $v$, by Taylor series expansions, we can derive the
discretization error as follows
\begin{eqnarray*}
\begin{array}{rcl}
&&\Big|(\psi\frac{\partial v}{\partial c})_i^{n+1}
-\sup_{\sigma\in\{\underline{\sigma},\overline{\sigma}\}}(L^{\sigma}v^{n+1})_i-
\left[ \frac{v_i^{n+1}-\bar{v}_i^{n}}{\Delta
t_{n+1}}-\sup_{\sigma^{n+1}\in\{\underline{\sigma},\overline{\sigma}\}}(L_h^{\sigma^{n+1}}v^{n+1})_i\right]\\
&\leq & |(\psi\frac{\partial v}{\partial
\tau})_i^{n+1}-\frac{v_i^{n+1}-\bar{v}_i^{n}}{\Delta
t_{n+1}}|+\sup_{\sigma\in\{\underline{\sigma},\overline{\sigma}\}}|(L^{\sigma}v^{n+1})_i-(L_h^{\sigma}v^{n+1})_i|\\
&=&O(\Delta t)+O(\Delta x)
\end{array}
\end{eqnarray*}
which prove the consistency of the discretizaiton scheme.
$\square$

\begin{lemma} (Monotonicity) The discretization ($\ref{eq610}$) is monotone.
\end{lemma}
{\bf Proof.} For $i=0$ or $i=N$ the lemma is trivially true. For
$0<i<N$, we write equation ($\ref{eq612}$) in component form
\begin{eqnarray*}
&&G_i^{n+1}(h,u_{i-1}^{n+1},u_{i}^{n+1},u_{i+1}^{n+1},\bar{u}_i^{n})\\
&=&
\frac{u_i^{n+1}-u_i^{n}}{\Delta t_{n+1}}-
\frac{\bar{u}_i^{n}-u_i^{n}}{\Delta
t_{n+1}}\\
&&-\sup_{\sigma_i^{n+1}\in\{\underline{\sigma},\overline{\sigma}\}}
[\alpha_i^{n+1}(\sigma_i^{n+1})u_{i-1}^{n+1}+\beta_i^{n+1}(\sigma_i^{n+1})u_{i+1}^{n+1}
-(\alpha_i^{n+1}(\sigma_i^{n+1})+\beta_i^{n+1}(\sigma_i^{n+1})+r)u_{i}^{n+1}].\nonumber
\end{eqnarray*}

For $\varepsilon \geq 0$, we have
\begin{eqnarray*}
\begin{array}{rcl}
&&G_i^{n+1}(h,u_{i-1}^{n+1},u_{i}^{n+1},u_{i+1}^{n+1}+\varepsilon,\bar{u}_i^{n})
-
G_i^{n+1}(h,u_{i-1}^{n+1},u_{i}^{n+1},u_{i+1}^{n+1},\bar{u}_i^{n})\\
&\leq&\sup_{\sigma_i^{n+1}\in\{\underline{\sigma},\overline{\sigma}\}}\{-\beta_i^{n+1}(\sigma^{n+1})\varepsilon\}=-\varepsilon
\inf_{\sigma_i^{n+1}\in\{\underline{\sigma},\overline{\sigma}\}}\{\beta_i^{n+1}(\sigma^{n+1})\}\leq
0.
\end{array}
\end{eqnarray*}
With the similar argument we derive
\begin{eqnarray*}
G_i^{n+1}(h,u_{i-1}^{n+1}+\varepsilon,u_{i}^{n+1},u_{i+1}^{n+1},\bar{u}_i^{n})
-
G_i^{n+1}(h,u_{i-1}^{n+1},u_{i}^{n+1},u_{i+1}^{n+1},\bar{u}_i^{n})\leq
0.
\end{eqnarray*}

It is easy to check that
\begin{eqnarray*}
G_i^{n+1}(h,u_{i-1}^{n+1},u_{i}^{n+1},u_{i+1}^{n+1},\bar{u}_i^{n}+\varepsilon)
-
G_i^{n+1}(h,u_{i-1}^{n+1},u_{i}^{n+1},u_{i+1}^{n+1},\bar{u}_i^{n})=-\frac{\varepsilon}{\Delta
t_{n+1}}\leq 0.
\end{eqnarray*}

Thus we proved the discretization is monotone. $\ \ \square$

The discrete scheme ($\ref{eq610}$) is consistent, stable and
monotone and from \cite{Barles} we have the following convergence
theorem
\begin{theorem} (Convergence to the viscosity solution)
The solution of the discrete scheme ($\ref{eq610}$) converges to
the viscosity solution of equation ($\ref{HJB11}$).
\end{theorem}

\subsection{Iterative Solution of  Discrete Algebraic System}

In the previous subsection, we show that the solution of the
discretization ($\ref{eq610}$) convergences to the viscosity
solution of the nonlinear PDE ($\ref{HJB11}$). Since the implicit
scheme leads a nonlinear algebraic system ($\ref{eq610}$) at each
timestep, the discretization is not a practical scheme. In this
section, we aim to solve this discrete scheme by a practical
iterative method.

Iterative Algorithm for ($\ref{eq610}$)

1. $n=0$

2. Set $k=0$ and $\tilde{u}^k=U^n$

3. For $k=0,1,2, \cdots$

Solve
\begin{eqnarray}\label{eq615}
\begin{array}{l}
[I-\Delta t_{n+1}A^{n+1}(\tilde{\sigma}^k)]\tilde{u}^{k+1}=
\bar{U}^{n}\\
\tilde{\sigma}_i^k\in
\mbox{argsup}_{\sigma\in\{\underline{\sigma},\overline{\sigma}\}}
\{[A^{n+1}(\sigma)]\tilde{u}^{k}]_i\}
\end{array}\label{IT}
\end{eqnarray}

4. If $\max_{i}\frac{|\tilde{u}_i^{k+1}-\tilde{u}_i^{k}|}{\max(scale,\tilde{u}_i^{k+1})}< tolerance$ then quit, else $k=k+1$ go to 3.

5. Set
$U^{n+1}=\tilde{u}^{k+1},\hat{\sigma}^{n+1}=\tilde{\sigma}^k$ and
$n=n+1$, go to 2.

The term $scale$ is used to ensure that unrealistic levels of accuracy are not required when the value is very small.
In the iterative algorithm $\tilde{\sigma}^k$ is given by
$$
\tilde{\sigma}_i^k=\left\{\begin{array}{ll}
\overline{\sigma},& \mbox{if} \frac{(\tilde{u}^k_{i+1}-\tilde{u}^k_{i})/(x_{i+1}-x_i)
-(\tilde{u}^k_{i}-\tilde{u}^k_{i-1})/(x_{i}-x_{i-1})}{x_{i+1}-x_{i-1}}\ge 0.\\
\underline{\sigma},& \mbox{if}
\frac{(\tilde{u}^k_{i+1}-\tilde{u}^k_{i})/(x_{i+1}-x_i)
-(\tilde{u}^k_{i}-\tilde{u}^k_{i-1})/(x_{i}-x_{i-1})}{x_{i+1}-x_{i-1}}<
0.
\end{array}\right.
$$

\begin{theorem}(Convergence of the Iterative Algorithm) The iteration algorithm ($\ref{eq615}$) for ($\ref{eq610}$)
convergences to the unique solution of equation ($\ref{eq67}$) for
any initial iterate $\tilde{u}^0$.
\end{theorem}
{\bf Proof.} First, we will prove that $\|\tilde{u}^k\|_{\infty}$
is bounded independent of iteration $k$ with a similar argument in
Lemma $\ref{le61}$.

We can write ($\ref{IT}$) in component form as follows

\begin{eqnarray*}
(1+\Delta
t_{n+1}(\alpha_i^{n+1}(\tilde{\sigma}^k)+\beta_i^{n+1}(\tilde{\sigma}^k)+r))\tilde{u}_{i}^{k+1}
-\Delta
t_{n+1}\alpha_i^{n+1}(\tilde{\sigma}^k)\tilde{u}_{i-1}^{k+1}
-\Delta
t_{n+1}\beta_i^{n+1}(\tilde{\sigma}^k)\tilde{u}_{i+1}^{k+1}
=\bar{u}_i^k.
\end{eqnarray*}
Then
\begin{eqnarray*}
&&(1+\Delta
t_{n+1}(\alpha_i^{n+1}(\tilde{\sigma}^k)+\beta_i^{n+1}(\tilde{\sigma}^k)+r))|\tilde{u}_{i}^{k+1}|\\
&\leq & |\tilde{u}^n|+|\Delta
t_{n+1}\alpha_i^{n+1}(\tilde{\sigma}^k)\tilde{u}_{i-1}^{k+1}
+\Delta
t_{n+1}\beta_i^{n+1}(\tilde{\sigma}^k)\tilde{u}_{i+1}^{k+1}|\\
&\leq &\|U^n\|_{\infty}+\Delta
t_{n+1}(\alpha_i^{n+1}(\tilde{\sigma}^k)+\beta_i^{n+1}(\tilde{\sigma}^k)\|\tilde{u}^{k+1}\|_{\infty}.
\end{eqnarray*}
Thus
\begin{eqnarray*}
\|\tilde{u}^{k+1}\|_{\infty}\leq \|U^n\|_{\infty} \leq
\max(\|U^0\|_{\infty},C_b)
\end{eqnarray*}
which means that $\|\tilde{u}^{k+1}\|_{\infty}$ is bounded
independent of $k$.

Now we will prove that the iterates $\{\tilde{u}^k\}$ form a nondecreasing sequence.
 From ($\ref{IT}$), the iterates difference $\tilde{u}^{k+1}-\tilde{u}^{k}$ satisfy

\begin{eqnarray}\label{eq616}
[I-\Delta
t_{n+1}A^{n+1}(\tilde{\sigma}^k)](\tilde{u}^{k+1}-\tilde{u}^{k})=\Delta
t_{n+1}[A^{n+1}(\tilde{\sigma}^k)-A^{n+1}(\tilde{\sigma}^{k-1})]\tilde{u}^{k}.
\end{eqnarray}
Notice that
$$
\tilde{\sigma}_i^k\in
\mbox{argsup}_{\sigma\in\{\underline{\sigma},\overline{\sigma}\}}
\{[A^{n+1}(\sigma)]\tilde{u}^{k}]_i\},
$$
the right side of ($\ref{eq616}$) is nonnegative, i.e.
\begin{eqnarray*}
\Delta
t_{n+1}\{[A^{n+1}(\tilde{\sigma}^k)-A^{n+1}(\tilde{\sigma}^{k-1})]\tilde{u}^{k}\}_i\ge
0
\end{eqnarray*}
Consequently,
\begin{eqnarray}\label{eq617}
[(I-\Delta
t_{n+1}A^{n+1}(\tilde{\sigma}^k))(\tilde{u}^{k+1}-\tilde{u}^{k})]_i\ge
0.
\end{eqnarray}
From Theorem $\ref{th61}$ we know that matrix $[I-\Delta
t_{n+1}A^{n+1}(\tilde{\sigma}^k)]$ is an M-matrix and hence
\begin{eqnarray}\label{eq618}
[I-\Delta t_{n+1}A^{n+1}(\tilde{\sigma}^k)]^{-1}\ge 0.
\end{eqnarray}

From ($\ref{eq617}$) ($\ref{eq618}$), we can derive that
\begin{eqnarray}
\tilde{u}^{k+1}-\tilde{u}^{k}\ge 0
\end{eqnarray}
which prove that the iterates form a nondecreasing sequence. The
iterates sequence $\{\tilde{u}^{k}\}$ is nondecreasing and
bounded, thus the sequence converges to a solution, i.e., $\exists
\tilde{u}, s.t.
\|\tilde{u}^{k}-\tilde{u}\|_{\infty}\longrightarrow 0$ and
\begin{eqnarray}\label{eq619}
[I-\Delta t_{n+1}A^{n+1}(\tilde{\sigma})]\tilde{u}= \bar{u}^{n},\\
\tilde{\sigma}_i\in
\mbox{argsup}_{\sigma\in\{\underline{\sigma},\overline{\sigma}\}}
\{[A^{n+1}(\sigma)]\tilde{u}]_i\}.\nonumber
\end{eqnarray}
Since the matrix $[I-\Delta t_{n+1}A^{n+1}(\tilde{\sigma}^k)]$ is
a M-matrix, thus the solution of ($\ref{eq619}$) is unique.
$\square$

In next section we will simulate the ask (resp. bid) price of the
contingent claim by using monotone characteristic finite
difference schemes for ($\ref{eq521}$) (resp. ($\ref{eq522}$)).
The convergence of the solution of the monotone characteristic
finite difference schemes ($\ref{eq610}$) to the viscosity
solution of ($\ref{HJB11}$) guarantee the simulation ask (resp.
bid) price of the contingent claim convergence to the correct
financial relevant solution.

\section{Examples and simulations}
In this section, we will give simulations for the bid-ask pricing
mechanisms of contingent claims under uncertainty with payoff
given by some function $\phi(S_T)$. In computational simulations,
we only make the numerical program for the nonlinear PDE
($\ref{eq521}$), since the bid price $u^b(t,x):=-u(t,x)$ where
$u(t,x)$ is the viscosity solution of ($\ref{eq521}$) with
terminal condition $u(T,x)=-\phi(x)$.


\begin{example}
Digital call option under uncertain volatility.
\end{example}
We consider a digital call option with the payoff as follows
\begin{eqnarray*}
\phi(S_T)=\left\{\begin{array}{ll}
1,&S_T\ge K\\
0,&S_T<K
\end{array}\right..
\end{eqnarray*}
The strike price of the digital option is $K=100(\$)$ and the
maturity is six months $T=0.5$ year. The volatility bounds are
given by $\underline{\sigma}=0.15,\overline{\sigma}=0.25$ and
short interest rate is $r=0.10$.

\begin{figure}[ht]
\includegraphics[scale=0.7]{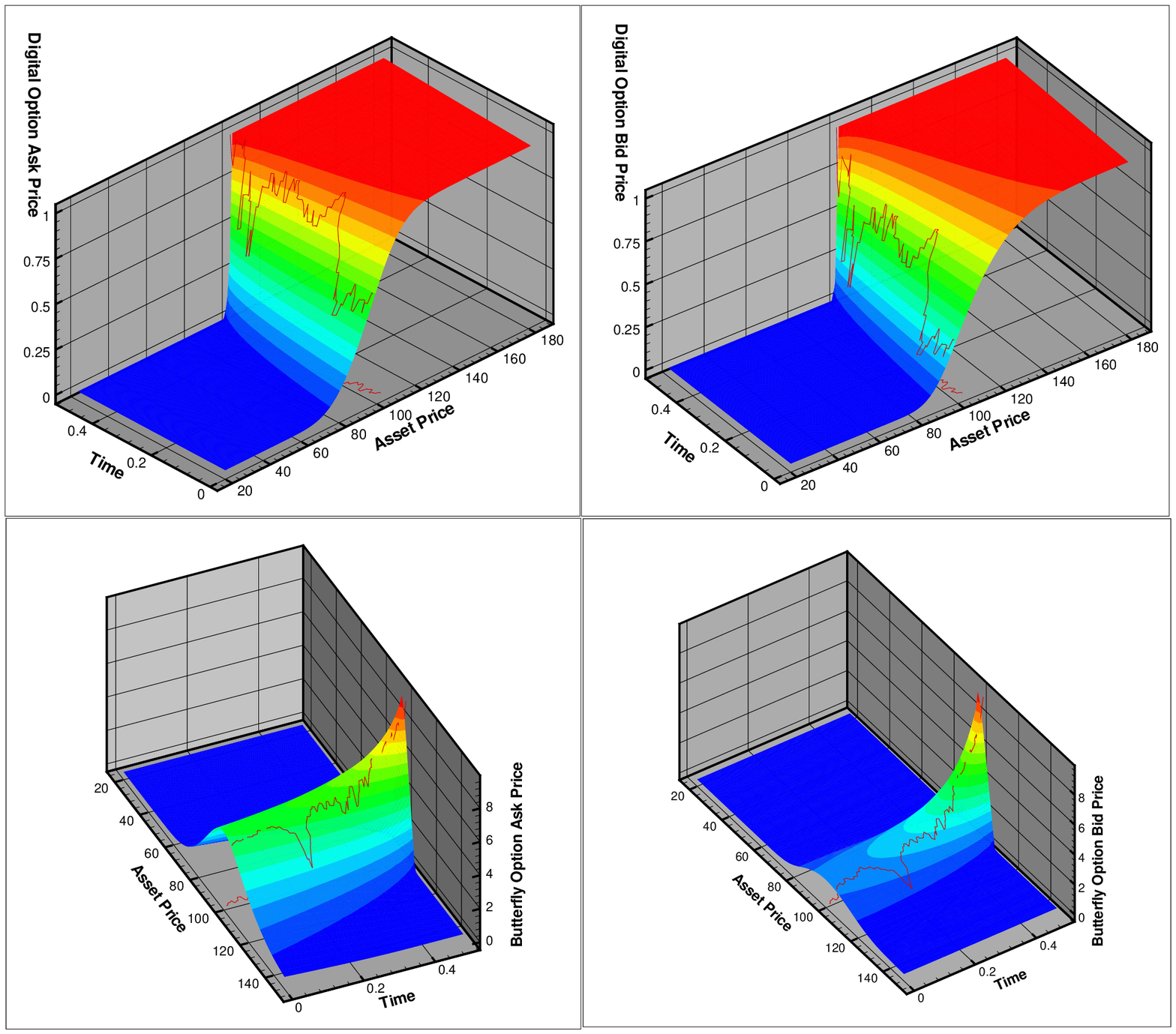}
\parbox{15cm}{\caption{\label{fg1-1}Price trajectory on the ask
(left top) and bid (right top) price surface of the digital option
(up) and butterfly option (down).}}
\end{figure}

We use the numerical schemes constructed in Section 6 to compute
the nonlinear PDE $(\ref{eq521})$ with the payoff function as
initial condition and the boundary condition is
$\phi(S_{\mbox{max}})=1$. We choose the grid as $\Delta s=1,
\Delta t =0.0025$ and iterative tolerance$=10^{-6}$. We plot Fig.
$\ref{fg1-1}$ the price trajectory on the ask (top left) price
 and bid (top right) price surfaces of the digital call
option.

\begin{example}
Butterfly option under uncertain volatility.
\end{example}
The second example is a butterfly option with the payoff as
follows
\begin{eqnarray*}
\phi(S_T)=\max(S-K_1,0)-2\max (S-(K_1+K_2)/2,0)+\max(S-K_2,0),
\end{eqnarray*}
the boundary condition is $u^a(S_{\mbox{max}})=0, K_1=90$ and
$K_2=110$. The other parameters used here are the same as that we
used in example 1. The ask price (down left) and the bid price
(down right) surfaces of the butterfly option are shown in Fig.
$\ref{fg1-1}$.

Fig. $\ref{fg1-1}$ shows that the ask price trajectory is above
the bid price trajectory for the both claims, and the ask (resp.
bid) price dynamic keeps the monotone intervals and convex (resp.
 concave) intervals of the corresponding payoff function which verify the theoretical results we showed in Lemma
 $\ref{le51}$.

\begin{thebibliography}{99}
\bibitem{Artzner2}
Artzner, Ph., Delbaen F., Eber J. M. (1999) Coherent measures of
risk, Mathematical Finance. 9, 73-88.

\bibitem{Avell}
Avellaneda, M., Levy, A. and Par$\acute{\mbox{a}}$s, A. (1995)
Pricing and Hedging Derivative Securities in Markets With
Uncertain Volatilities, Appl. Math. Finance. 2, 73-88.

\bibitem{Baren1}
Barenblatt, G.I. (1978) Similarity, self-similarity and
intermediate asympototics, Consultants Bureau, New York (there
exists a revised second Russian edition, Leningrad
Gidrometeoizdat, 1982).

\bibitem{Baren2}
Barenblatt, G.I. and Sivashinski, G.I. (1969) Self-similar
solutions of the second kind in nonlinear filtration, Appl. Math.
Mech. 33, 836-845(translated from Russian PMM, pages 861-870).

\bibitem{Barles}
Barles, G. and Jakobsen, E.R. (2007) Error bounds for monotone
approximation schemes for parabolic Hamilton-Jacobi-Bellman
equations, Mathematics of Computation, 76, 1861-1893.

\bibitem{Bion1}
Bion-Nadal, J. (2009) Time consistent dynamic risk processes,
Stochastic Processes and Their Applications, 119, 633-654.

\bibitem{Bion2}
Bion-Nadal, J. (2009) Bid-ask dynamic pricing in financial markets
with transaction costs and liquidity risk, Journal of Mathematical
Economics, 45, 738-750.

\bibitem{Chaumont}
Chaumont, S. (2003) A strong comparision result for viscosity solutions to Hamilton-Jacobi-Bellman equations with Dirichlet condition on a non-smooth boundary, Acad. Sci. Paris, Ser. I 336.

\bibitem{ChernyMadan}
Cherny, A., Madan, D.B. (2009) New measures for performance
evaluation, The Review of Financial Studies, 22/7, 2571-2606.

\bibitem{Cherny}
Cherny, A., Madan, D.B. (2010) Illiquid market as a counterparty:
an introduction to conic finance, International Journal of
Theoretical and Applied Finance, 13/8, 1449-1177.

\bibitem{Lions}
Crandall, M.G., Ishii, H., Lions, P.L. (1992) User's guide to
viscosity solutions of second order partial differential
equations, Bull. Amer. Math. Soc. 27(1), 1-67.

\bibitem{Douglas}
Douglas, J.Jr., Russell, T.F. (1982) Numerical method for
convection-dominated diffusion problem based on combing the method
of characteristics with finite element or finite difference
procedures, SIAM J. Numer. Anal., 19(5), 871-885.


\bibitem{Denis}
Denis, L., Hu, M. and Peng, S. (2010) Function spaces and capacity
related to a Sublinear Expectation: application to G-Brownian
Motion Paths, Potential Analysis, 34, 139-161.




\bibitem{Eberlein}
Eberlein, E., Madan, D.B., Pistorius, M., Schouotens, W., Yor, M.
(2012) Two price economies in continuous time, Preprint (August,
2012),
http://www.stochastik.uni-freiburg.de/eberlein/papers/TPECT1.pdf.

\bibitem{Epstein}
Epstein, L.,  Ji, Shaolin. (2011) Ambiguous volatility,
possibility and utility in continuous time, arXiv:1103.1652v4.

\bibitem{Fleming}
Fleming, W., Soner, M. (1992) {\it Controlled markov processes and
viscosity solutions}. Springer Verlag, New York.

\bibitem{Lyons}
Lyons, T. J. (1995) Uncertain volatility and the risk-free
synthesis of derivatives, Appl. Math. Finance, 2, 117-133.

\bibitem{Madan}
Madan, D.B. (2012) A two price theory of financial equilibrium
with risk management implications, Ann Finance, 8, 489-505.



\bibitem{PengKaroui}
El Karoui, Peng, S., Quenez, M.-C. (1997) Backward stochastic
differential equations in finance, Math. Finance. 7, 1-71.




\bibitem{PengA}
Peng, S. (1992) A generalized dynamic programming principle and
Hamilton-Jacobi-Bellman equation. Stochastics and Stochastic
Reports, 38(2): 119-134.

\bibitem{Peng3}
Peng, S., (2004) Filtration Consistent Nonliear Expectations and
Evaluations of Contingent Claims. Acta Mathematicae Applicatae
Sinica, English Series, 20(2), 1-24.

\bibitem{PengB}
Peng, S. (2005) G-Expectation, G-Brownian Motion and Related
Stochastic Calculus of Ito Type, Stochastic Analysis and
Applications, The Abel Symposium, 541-567.

\bibitem{PengD}
Peng, S. (2008) Multi-dimensional G-Brownian Motion and Related
stochastic Calculus under G-Expectation, Stochastic Processes and
their Applications, 118, 2223-2253.

\bibitem{PengC}
Peng, S. (2010) Nonlinear expectations and stochastic calculus
under uncertainty - with robust central limit theorem and
G-Brownian Motion, Preprint arXiv:1002.4546v1.





\bibitem{Song}
Song, Y. (2011) Some properties on G-evaluation and its
applications to G-martingale decomposition, Science China
Mathematics, 54(2), 287-300.

\bibitem{Varga}
Varga, R.S. (2000) Matrix Iterative Analysis. Springer Verlag.
\end{thebibliography}
\end{document}